\documentclass[pra,showpacs,nofootinbib]{revtex4}
\usepackage{graphicx}

\def\be{\begin{equation}}
\def\ee{\end{equation}}
\def\ba{\begin{eqnarray}}
\def\ea{\end{eqnarray}}

\begin{document}
\title{Density-functional-theory calculations of matter in strong magnetic fields. I. Atoms and molecules}
\author{Zach Medin and Dong Lai}
\affiliation{Center for Radiophysics and Space Research, Department of Astronomy, Cornell University, Ithaca, New York 14853, USA}

\received{9 June 2006; published 14 December 2006}

\begin{abstract}
We present calculations of the electronic structure of various atoms
and molecules in strong magnetic fields ranging from $B=10^{12}$~G to
$2\times10^{15}$~G, appropriate for radio pulsars and magnetars. For
these field strengths, the magnetic forces on the electrons dominate
over the Coulomb forces, and to a good approximation the electrons are
confined to the ground Landau level.  Our calculations are based on
the density functional theory, and use a local magnetic
exchange-correlation function which is tested to be reliable in the
strong field regime. Numerical results of the ground-state energies
are given for H$_N$ (up to $N=10$), He$_N$ (up to $N=8$), C$_N$ (up to
$N=5$), and Fe$_N$ (up to $N=3$), as well as for various ionized atoms.
Fitting formulae for the $B$-dependence of the energies are also
given.  In general, as $N$ increases, the binding energy per atom in a
molecule, $|E_N|/N$, increases and approaches a constant value. For
all the field strengths considered in this paper, hydrogen, helium,
and carbon molecules are found to be bound relative to individual
atoms (although for $B$ less than a few $\times 10^{12}$~G, carbon
molecules are very weakly bound relative to individual atoms). Iron
molecules are not bound at $B\alt 10^{13}$~G, but become energetically
more favorable than individual atoms at larger field strengths.
\end{abstract}


\maketitle

\section{Introduction}

Neutron stars (NSs) are endowed with magnetic fields far beyond the reach of
terrestrial laboratories \citep{meszaros92,reisenegger05,harding06}.
Most of the $\sim 1600$ known radio pulsars
have surface magnetic fields in the range of $10^{11}-10^{13}$~G, as
inferred from their measured spin periods and period derivatives and
the assumption that the spindown is due to magnetic dipole
radiation. A smaller population of older, millisecond pulsars have
$B\sim 10^8-10^9$~G\@. For about a dozen accreting neutron stars in
binary systems, electron cyclotron features have been detected,
implying surface fields of $B\sim 10^{12}-10^{13}$~G\@.
An important development in
astrophysics in the last decade centered on the so-called anomalous
x-ray pulsars and soft gamma repeaters \citep{woods05}: there has been
mounting observational evidence in recent years that supports the idea
that these are magnetars, neutron stars whose radiations are
powered by superstrong magnetic fields of order 
$10^{14}-10^{15}$~G or higher \citep{duncan92,thompson95,thompson96}.
By contrast, the highest static magnetic field currently 
produced in a terrestrial laboratory is $5\times 10^5$~G;
transient fields approaching $10^9$~G have recently been generated during 
high-intensity laser interactions with dense plasmas \citep{wagner04}.

It is well-known that
the properties of matter can be drastically modified by strong magnetic
fields found on neutron star surfaces. The natural atomic unit for the
magnetic field strength, $B_0$, is set by equating the electron
cyclotron energy $\hbar\omega_{Be}=\hbar
(eB/m_ec)=11.577\,B_{12}$~keV, where $B_{12}=B/(10^{12}~{\rm G})$, to
the characteristic atomic energy $e^2/a_0=2\times 13.6$~eV (where
$a_0$ is the Bohr radius):
\be
B_0={m_e^2e^3c\over\hbar^3}=2.3505\times 10^9\, {\rm G}.
\label{eqb0}
\ee
For $b=B/B_0\agt 1$, the usual perturbative treatment of the magnetic
effects on matter (e.g., Zeeman splitting of atomic energy levels)
does not apply. Instead, in the transverse direction (perpendicular to
the field) the Coulomb forces act as a perturbation to the magnetic
forces, and the electrons in an atom settle into the ground Landau
level. Because of the extreme confinement of the electrons in the
transverse direction, the Coulomb force becomes much more effective in
binding the electrons along the magnetic field direction. The atom
attains a cylindrical structure. Moreover, it is possible for these
elongated atoms to form molecular chains by covalent bonding along the
field direction. Interactions between the linear chains can then lead
to the formation of three-dimensional condensed matter
\citep{ruderman74,ruder94,lai01}.

Our main motivation for studying matter in such strong magnetic fields
arises from the importance of understanding neutron star surface
layers, which play a key role in many neutron star processes and
observed phenomena.  Theoretical models of pulsar and magnetar
magnetospheres depend on the cohesive properties of the surface matter
in strong magnetic 
fields \citep{ruderman75,arons79,usov96,harding98,beloborodov06}.  
For example, depending on the cohesive energy of the surface matter, an
acceleration zone (``polar gap'') above the polar cap of a pulsar may
or may not form. More importantly, the surface layer directly mediates
the thermal radiations from neutron stars. The advent of x-ray
telescopes in recent years has made detailed study of neutron star
surface emission a reality. Such study can potentially provide
invaluable information on the physical properties and evolution of NSs:
equation of state at supernuclear densities, superfluidity, cooling
history, magnetic field, surface composition, different NS
populations, etc. (see, e.g., Ref.~\citep{yakovlev04}). More than two
dozen isolated neutron stars (including radio pulsars, radio-quiet
neutron stars and magnetars) have clearly detected thermal surface emission
\citep{kaspi05,haberl05,harding06}. While
some neutron stars show featureless spectra, absorption lines or features
have been detected in half a dozen or so systems \citep{haberl05}.
Indeed, many of the observed neutron stars have sufficiently low
surface temperatures and/or high magnetic fields, such that bound
atoms or molecules are expected to be present in their
atmospheres \citep{lai97,potekhin99,ho03,potekhin04}.  
It is even possible that the atmosphere is condensed into a solid or 
liquid form from which radiation directly emerges
\citep{lai97,vanadelsberg05,lai01}.
Thus, in order to properly interpret various observations of neutron stars,
it is crucial to have a detailed understanding of the properties of 
atoms, molecules and condensed matter in 
strong magnetic fields ($B\sim 10^{11}$-$10^{16}$~G).

\subsection{Previous works}

H and He atoms at almost all field strengths have been well studied
\citep{ruder94,jones99,alhujaj03}, including the nontrivial effect
associated with the center-of-mass motion of a H atom
\citep{potekhin98}. \citet{neuhauser87} presented numerical results
for several atoms up to $Z=26$ (Fe) at $B\sim 10^{12}$~G based on
calculations using a one-dimensional Hartree-Fock method (see also
Ref.~\citep{mori02} for $Z$ up to 10).  Some results [based on a
two-dimensional (2D) mesh Hartree-Fock method] for atoms (up to
$Z=10$) at the field strengths $B/B_0=0.5-10^4$ are also available
\citep{ivanov00,alhujaj04a, alhujaj04b}. The Hartree-Fock method is
approximate because electron correlations are neglected. Due to their
mutual repulsion, any pair of electrons tend to be more distant from
each other than the Hartree-Fock wave function would indicate. In
zero-field, this correlation effect is especially pronounced for the
spin-singlet states of electrons for which the spatial wave function
is symmetrical. In strong magnetic fields ($B\gg B_0$), the electron
spins (in the ground state) are all aligned antiparallel to the
magnetic field, and the multielectron spatial wave function is
antisymmetric with respect to the interchange of two electrons. Thus
the error in the Hartree-Fock approach is expected to be less than the
$1\%$ accuracy characteristic of zero-field Hartree-Fock calculations
\citep{neuhauser87,schmelcher99}. Other calculations of heavy atoms in
strong magnetic fields include Thomas-Fermi type statistical models
\citep{fushiki92,lieb94a,lieb94b} and density functional theory
\citep{jones85,jones86, kossl88,relovsky96}. The Thomas-Fermi type
models are useful in establishing asymptotic scaling relations, but
are not adequate for obtaining accurate binding and excitation
energies. The density functional theory can potentially give results
as accurate as the Hartree-Fock method after proper calibration is
made \citep{vignale87,vignale88}.

Quantitative results for the energies of hydrogen molecules H$_N$ with
$N=2,3,4,5$ in a wide range of field strengths ($B\gg B_0$) were
obtained (based on the Hartree-Fock method) by 
Lai {\it et al.}~\citep{lai92,lai01} and molecular excitations were studied in
Refs.~\citep{lai96,schmelcher01} (more complete references can be found in
Ref.~\citep{lai01}). Quantum Monte Carlo
calculations of H$_2$ in strong magnetic fields have been performed
\citep{ortiz95}. Some numerical results of He$_2$ for various field strengths
are also
available \citep{lai01}. Hartree-Fock results of diatomic
molecules (from H$_2$ up to C$_2$) and several larger molecules
(up to H$_5$ and He$_4$) at $B/B_0=1000$ are given in Ref.~\citep{demeur94}.

\subsection{Plan of this paper}

In this paper and its companion paper \citep{medin06a}, we develop a
density-functional-theory calculation of the ground-state energy
of matter for a wide range of magnetic field strengths, from $10^{12}$~G
(typical of radio pulsars) to $2\times 10^{15}$~G (magnetar fields).
We consider H, He, C, and Fe, which represent the most likely composition
of the outermost layer of neutron stars (e.g., Ref.~\citep{harding06}).
The present paper focuses on atoms
(and related ions) and small molecules. Because of additional complications
related to the treatment of band structure, calculations of
infinite molecular chains and condensed
matter are presented in Ref.~\citep{medin06a}.

Our calculations are based on density functional theory
\citep{hohen64,kohn65,jones89}. As mentioned above, the Hartree-Fock
method is expected to be highly accurate, particularly in the strong
field regime where the electron spins are aligned with each other. In
this regime the density functional method is not as accurate, due to
the lack of an exact correlation function for electrons in strong
magnetic fields. However, in dealing with systems with many
electrons, the Hartree-Fock method becomes increasingly
impractical as the magnetic field increases, since more and more
Landau orbitals (even though electrons remain in the ground Landau
level) are occupied and keeping track of the direct and exchange
interactions between electrons in various orbitals becomes
computationally rather tedious. Our density-functional calculations
allow us to obtain the energies of atoms and small molecules and the
energy of condensed matter using the same method, thus providing
reliable cohesive energy of condensed surface of magnetic neutron
stars, a main goal of our study. Compared to previous
density-functional-theory calculations \citep{jones85,jones86,
kossl88,relovsky96}, we use an improved exchange-correlation function
for highly magnetized electron gases, we calibrate our density
functional code with previous results (when available) based on other
methods, and (for calculations of condensed matter) adopt a more
accurate treatment of the band structure. Moreover, our calculations
extend to the magnetar-like field regime ($B\sim 10^{15}$~G).

Note that in this paper we neglect the motions of the nuclei, due to
electron-nucleus interactions or finite temperatures. The
center-of-mass motions of the atoms and molecules induce the motional
Stark effect, which can change the internal structure of the bound
states (see, e.g., Refs.~\citep{lai01,potekhin98}). Such issues are
beyond the scope of this paper.

After summarizing the approximate scaling relations for atoms and
molecules in strong magnetic fields in Sec.~II, we describe our method
in Sec.~III and present numerical results in Sec.~IV\@. Some technical
details are given in the Appendix.

\section{Basic scaling relations for atoms and molecules in strong 
magnetic fields}

\subsection{Atoms}

First consider a hydrogenic atom (with one electron and nuclear charge
$Z$).  In a strong magnetic field with $b=B/B_0\gg Z^2$, the electron
is confined to the ground Landau level (``adiabatic approximation''),
and the Coulomb potential can be treated as a perturbation. The energy
spectrum is specified by two quantum numbers, $(m,\nu)$, where
$m=0,1,2,\ldots$ measures the mean transverse separation between the
electron and the nucleus ($-m$ is also known as the magnetic quantum
number), while $\nu$ specifies the number of nodes in the
$z$ wave function.  There are two distinct types of states in the
energy spectrum $E_{m\nu}$.  The ``tightly bound'' states have no node
in their $z$ wave functions ($\nu=0$). The transverse size of the atom
in the $(m,0)$ state is $L_\perp\sim\rho_m=(2m+1)^{1/2}\rho_0$, with
$\rho_0=(\hbar c /eB)^{1/2}=b^{-1/2}$~(in atomic
units).\footnote{Unless otherwise specified, we use atomic units, in
which length is in $a_0$ (Bohr radius), mass in $m_e$, energy in
$e^2/a_0=2$~Ry, and field strength in units of $B_0$.}  For $\rho_m\ll
1$, the atom is elongated with $L_z\gg L_\perp$.  We can estimate the
longitudinal size $L_z$ by minimizing the energy, $E\sim L_z^{-2}-Z
L_z^{-1}\ln (L_z/L_\perp)$ (where the first term is the kinetic energy
and the second term is the Coulomb energy), giving
\be
L_z\sim \left(Z\ln {1\over Z\rho_m}\right)^{-1}.
\ee
The energy is given by
\be
E_{m0}\sim -Z^2\,\left[\ln{1\over Z^2}\left({b\over 2m+1}\right)
\right]^2
\label{eqem3}\ee
for $b\gg (2m+1)Z^2$. Another type of state of the atom has nodes in 
the $z$ wave functions ($\nu>0$). These states are ``weakly bound'', 
and have energies given by $E_{m\nu}\simeq -Z^2n^{-2}$~Ry, where  $n$ is the 
integer part of $(\nu+1)/2$. The sizes of the wave functions are $\rho_m$
perpendicular to the field and $L_z\sim \nu^2/Z$ along the field
(see Ref.~\citep{lai01} and references therein for more details).

A multielectron atom (with the number of 
electrons $N_e$ and the charge of the nucleus $Z$) can be constructed 
by placing electrons at the lowest available energy levels of a hydrogenic 
atom. The lowest levels to be filled are the tightly bound states with 
$\nu=0$. When $a_0/Z \gg \sqrt {2 N_e-1}\rho_0$, i.e., $b \gg 2 Z^2N_e$,
all electrons settle into the tightly bound levels with 
$m=0,1,2,\cdots,N_e-1$. The energy of the atom is approximately given
by the sum of all the eigenvalues of Eq.~(\ref{eqem3}). Accordingly, we 
obtain an asymptotic expression for $N_e \gg 1$ \citep{kadom71}:
\be
E \sim - Z^2N_e\left(\ln{b\over 2Z^2N_e}\right)^2.
\ee

For intermediate-strong fields (but still strong enough to ignore
Landau excitations), $Z^2N_e^{-2/3} \ll b \ll 2 Z^2N_e$, many $\nu>0$
states of the inner Landau orbitals (states with relatively small $m$)
are populated by the electrons. In this regime a Thomas-Fermi type
model for the atom is appropriate, i.e., the electrons can be treated
as a one-dimensional Fermi gas in a more or less spherical atomic cell
(see, e.g., Refs.~\citep{kadom70,mueller71}). The electrons occupy the
ground Landau level, with the $z$ momentum up to the Fermi momentum
$p_F\sim n/b$, where $n$ is the number density of electrons inside the
atom (recall that the degeneracy of a Landau level is $e B /hc \sim
b$). The kinetic energy of electrons per unit volume is $\varepsilon_k
\sim b\,p_F^3\sim n^3/b^2$, and the total kinetic energy is $E_k \sim
R^3 n^3 /b^2 \sim N_e^3 /(b^2 R^6)$, where $R$ is the radius of the
atom. The potential energy is $E_p \sim -ZN_e/R$ (for $N_e\alt
Z$). Therefore the total energy of the atom can be written as $E \sim
{N_e^3/(b^2 R^6)} - {ZN_e/R}$. Minimizing $E$ with respect to $R$
yields
\be
R \sim (N_e^2/Z)^{1/5}b^{-2/5},\quad E \sim -(Z^2N_e)^{3/5}b^{2/5}.
\label{heavyatom}
\ee
For these relations to be valid, the electrons
must stay in the ground Landau level; this requires
$Z/R\ll\hbar\omega_{Be}=b$, which corresponds to $b\gg Z^2N_e^{-2/3}$.

\subsection{Molecules}

In a strong magnetic field, the mechanism of forming
molecules is quite different from the zero-field case \citep{ruderman74,lai92}.
Consider hydrogen as an example.
The spin of the electron in a H atom is aligned antiparallel to the 
magnetic field (flipping the spin would cost $\hbar\omega_{Be}$),
therefore two H atoms in their ground states ($m=0$) 
do not bind together according to the exclusion principle. 
Instead, one H atom has to be excited to the $m=1$ state. The
two H atoms, one in the ground state ($m=0$), another
in the $m=1$ state then form the ground state of the H$_2$ molecule by
covalent bonding. Since the activation energy for exciting
an electron in the H atom from the Landau orbital $m$ to $(m+1)$
is small, the resulting H$_2$ molecule is stable.
Similarly, more atoms can be added to form H$_3$, H$_4,~\ldots$.
The size of the H$_2$ molecule is comparable to that of the H atom. 
The interatomic separation $a$ and the dissociation energy 
$D$ of the H$_2$ molecule scale approximately as 
$a\sim (\ln b)^{-1}$ and $D\sim (\ln b)^2$,
although $D$ is numerically smaller than the ionization energy of the H atom.

Consider the molecule Z$_N$, formed out of $N$ neutral atoms $Z$ (each with
$Z$ electrons and nuclear charge $Z$). For sufficiently large $b$ (see below),
the electrons occupy the Landau orbitals with $m=0,~1,~2,\ldots,NZ-1$, 
and the transverse size of the molecule is 
$L_\perp\sim (NZ/b)^{1/2}$. Let $a$ be the atomic spacing and $L_z\sim
Na$ the size of the molecule in the $z$ direction. 
The energy per ``atom'' in the molecule, $E=E_N/N$, can be written as
$E\sim Z (Na)^{-2}-(Z^2/a) l$, where $l\sim\ln (a/L_\perp)$. 
Variation of $E$ with respect to $a$ gives 
\be
a\sim (ZN^2l)^{-1},\qquad
E\sim -Z^3N^2l^2,\qquad {\rm with}~~l\sim \ln\left({b\over N^5Z^3}\right).
\ee
This above scaling behavior is valid for $1\ll N\ll N_s$.
The ``critical saturation number'' $N_s$ is reached when $a\sim L_\perp$,
or when
\citep{lai92}
\be
N_s\sim \left({b\over Z^3}\right)^{1/5}.
\ee
Beyond $N_s$, it becomes energetically more favorable
for the electrons to settle into the inner Landau orbitals (with smaller $m$)
with nodes in their longitudinal wave functions (i.e., $\nu\neq 0$). 
For $N\agt N_s$, the energy per atom
asymptotes to a value $E\sim -Z^{9/5}b^{2/5}$, 
and size of the atom scales as $L_\perp\sim a\sim Z^{1/5}b^{-2/5}$, 
independent of $N$ --- the molecule essentially becomes one-dimensional
condensed matter.

The scaling relations derived above are obviously crude --- they are
expected to be valid only in the asymptotic limit, 
$\ln(b/Z^3)\gg 1$. For realistic neutron stars, this limit is 
not quite reached. Thus these scaling results should only serve 
as a guide to the energies of various molecules. For a given field
strength, it is not clear from the above analysis whether the Z$_N$
molecule is bound relative to individual atoms. To answer this
question requires quantitative calculations.

\section{Density-functional calculations: Methods and equations}

Our calculations will be based on the ``adiabatic approximation,'' in
which all electrons are assumed to lie in the ground Landau level.
For atoms or molecules with nucleus charge number $Z$, this is an
excellent approximation for $b\gg Z^2$. Even under more relaxed
condition, $b\gg Z^{4/3}$ (assuming the number of electrons in each
atom is $N_e\sim Z$) this approximation is expected to yield a
reasonable total energy of the system and accurate results for the
energy difference between different atoms and molecules; a
quantitative evaluation of this approximation in this regime is beyond
the scope of this paper (but see
Refs.~\citep{ivanov00,alhujaj04a,alhujaj04b}).

In the adiabatic approximation, the one-electron wave function (``orbital'')
can be separated into a transverse (perpendicular to the external 
magnetic field) component and a longitudinal (along the magnetic field)
component:
\be
\Psi_{m\nu}(\mathbf{r}) = 
W_m(\mathbf{r_\perp})f_{m\nu}(z)\,.  
\ee
Here $W_m$ is the ground-state Landau wave function \citep{landau77}
given by 
\be W_m(\mathbf{r_\perp}) =
\frac{1}{\rho_0 \sqrt{2\pi m!}}
\left(\frac{\rho}{\sqrt{2}\rho_0}\right)^m
\exp\left(\frac{-\rho^2}{4\rho_0^2}\right) \exp(-im\phi) \,,
\label{Wmeq}
\ee
where $\rho_0=(\hbar c/eB)^{1/2}$ is the cyclotron radius (or magnetic length),
and $f_{m\nu}$ is the longitudinal wave function which must be solved 
numerically. We normalize $f_{m\nu}$ over all space:
\be
\int_{-\infty}^\infty dz \, |f_{m\nu}(z)|^2 = 1 \,,
\ee
so that $\int d\mathbf{r} \, |\Psi_{m\nu}(\mathbf{r})|^2 = 1$.
The density distribution of electrons in
the atom or molecule is
\be 
n(\mathbf{r}) = \sum_{m\nu}
|\Psi_{m\nu}(\mathbf{r})|^2 = \sum_{m\nu} |f_{m\nu}(z)|^2 |W_m|^2(\rho) \,,
\label{densityeq}
\ee
where the sum is over all the electrons in the atom or molecule,
with each electron occupying an $(m\nu)$ orbital. The notation
$|W_m|^2(\rho)=|W_m(\mathbf{r}_\perp)|^2$ is used here because $W_m$
is a function of $\rho$ and $\phi$ but $|W_m|^2$ is a function only of
$\rho$.

In an external magnetic field, the Hamiltonian of a free electron is
\be
\hat{H} = \frac{1}{2m_e}\left(\mathbf{p}+\frac{e}{c}\mathbf{A}\right)^2+\frac{\hbar eB}{2m_e c}\sigma_z \,,
\ee
where $\mathbf{A}=\frac{1}{2}\mathbf{B}\times\mathbf{r}$ is the vector potential of the external magnetic field and $\sigma_z$ is the $z$ component Pauli spin matrix. For electrons in Landau levels, with their spins aligned parallel/antiparallel to the magnetic field, the Hamiltonian becomes
\be
\hat{H} = \frac{\hat{p}_z^2}{2m_e}+\left(n_L+\frac{1}{2}\right)\hbar\omega_{Be}\pm\frac{1}{2}\hbar\omega_{Be} \,,
\ee
where $n_L=0,1,2,\cdots$ is the Landau level index; for electrons in the ground Landau level, with their spins aligned antiparallel to the magnetic field (so $n_L=0$ and $\sigma_z \rightarrow -1$),
\be
\hat{H} = \frac{\hat{p}_z^2}{2m_e} \,.
\ee
The total Hamiltonian for the atom or molecule then becomes
\be
\hat{H} = \sum_i \frac{\hat{p}_{z,i}^2}{2m_e} + V \,,
\ee
where the sum is over all electrons and $V$ is the total potential energy of the atom or molecule. From this we can derive the total energy of the system.

Note that we use nonrelativistic quantum mechanics in our calculations, 
even when $\hbar\omega_{Be}\agt m_ec^2$ or 
$B\agt B_Q=B_0/\alpha^2=4.414\times 10^{13}$~G\@. 
This is valid for two reasons: 
(i) The free-electron energy in relativistic theory is 
\be
E=\left[c^2p_z^2+m_e^2c^4\left(1+2n_L {B\over B_Q}\right)\right]^{1/2}.
\label{eqrel}
\ee
For electrons in the ground Landau level ($n_L=0$), Eq.~(\ref{eqrel})
reduces to $E\simeq m_ec^2+p_z^2/(2m_e)$ for $p_zc\ll m_ec^2$; the
electron remains nonrelativistic in the $z$ direction as long as the
electron energy is much less than $m_ec^2$; (ii) Eq.~(\ref{Wmeq})
indicates that the shape of Landau transverse wave function is
independent of particle mass, and thus Eq.~(\ref{Wmeq}) is valid in
the relativistic theory.  Our calculations assume that the
longitudinal motion of the electron is nonrelativistic. This is valid
at all field strengths and for all elements considered with the
exception of iron at $B\agt 10^{15}$~G\@. Even at $B =
2\times10^{15}$~G (the highest field considered in this paper),
however, we find that the most-bound electron in any Fe atom or
molecule has a longitudinal kinetic energy of only $\sim 0.2m_ec^2$
and only the three most-bound electrons have longitudinal kinetic
energies $\agt 0.1m_ec^2$.  Thus relativistic corrections are small in
the field strengths considered in this paper. Moreover, we expect our
results for the relative energies between Fe atoms and molecules to be
much more accurate than the absolute energies of either the atoms or
the molecules.

Consider the molecule Z$_N$, consisting of $N$ atoms, each with 
an ion of charge $Z$ and $Z$ electrons. 
In the lowest-energy state of the system, the ions are aligned along 
the magnetic field. The spacing between ions, $a$, is
chosen to be constant across the molecule.
In the density functional theory, the total energy of the system 
can be represented as a functional of the total electron density 
$n(\mathbf{r})$:
\be
E[n] = E_K[n] + E_{eZ}[n] + E_{\rm dir}[n] + E_{\rm exc}[n] + E_{ZZ}[n] \,.
\ee
Here $E_K[n]$ is the kinetic energy of a system of noninteracting electrons,
and $E_{eZ}$, $E_{\rm dir}$, and $E_{ZZ}$ are the electron-ion Coulomb energy,
the direct electron-electron interaction 
energy, and the ion-ion interaction energy, respectively,
\ba
&&E_{eZ}[n] = -\!\!\sum_{j=1}^N Ze^2 \int d\mathbf{r} \,
\frac{n(\mathbf{r})}{|\mathbf{r} - \mathbf{z}_j|} \,, \\
&&E_{\rm dir}[n] = \frac{e^2}{2} \int \!\! \int d\mathbf{r}\,d\mathbf{r}' \,
\frac{n(\mathbf{r}) n(\mathbf{r}')}{|\mathbf{r} - \mathbf{r}'|} \,, \\
&&E_{ZZ}[n] = \sum_{j=1}^{N-1} (N-j)\frac{Z^2e^2}{ja} \,.
\ea
The location of the ions in the above equations is represented by
the set $\{\mathbf{z}_j\}$, with
\be
\mathbf{z}_j = (2j-N-1)\frac{a}{2}\hat{\mathbf{z}} \,.
\ee
The term $E_{\rm exc}$ represents exchange-correlation energy. 
In the local approximation, 
\be
E_{\rm exc}[n] = \int \! d\mathbf{r} \, n(\mathbf{r})\, 
\varepsilon_{\rm exc}(n) \,,
\ee
where $\varepsilon_{\rm exc}(n)=\varepsilon_{\rm ex}(n)
+\varepsilon_{\rm corr}(n)$ is the exchange and correlation
energy per electron in a uniform electron gas of density $n$. For
electrons in the ground Landau level, the (Hartree-Fock) exchange
energy can be written as follows \citep{danz71}:
\be 
\varepsilon_{\rm ex}(n) = -\pi e^2 \rho_0^2
n F(t) \,, 
\ee 
where the dimensionless function $F(t)$ is 
\be
F(t)=4\int_0^\infty\!\!\!dx\left[\tan^{-1}\left({1\over
x}\right)-{x\over 2}\ln\left(1+{1\over x^2}\right)\right] e^{-4tx^2},
\ee
and 
\be 
t =\left(\frac{n}{n_B}\right)^2 = 2\pi^4 \rho_0^6 n^2,
\ee
[$n_B=(\sqrt{2}\pi^2\rho_0^3)^{-1}$ is the density above which 
the higher Landau levels start to be filled in a uniform electron gas].
For small $t$, $F(t)$ can be expanded as follows \citep{fush89}:
\be 
F(t) \simeq 3-\gamma-\ln 4t +
\frac{2t}{3}\left(\frac{13}{6}-\gamma-\ln 4t \right) +
\frac{8t^2}{15}\left(\frac{67}{30}-\gamma-\ln 4t \right)
+{\cal O}(t^3\ln t),
\ee
where $\gamma= 0.5772\cdots$ is Euler's constant. We have found that
the condition $t\ll 1$ is well satisfied everywhere for almost all
molecules in our calculations. The notable exceptions are the carbon
molecules at $B=10^{12}$~G and the iron molecules at $B=10^{13}$~G,
which have $t\alt 1$ near the center of the molecule. These molecules
are expected to have higher $t$ values than the other molecules in our
calculations, as they have large $Z$ and low $B$.\footnote{For the
uniform gas model, $t \propto Z^{6/5} N_e^{-2/5} B^{-3/5}$.}

The correlation energy of uniform electron gas in strong magnetic
fields has not be calculated in general, except in the regime $t\ll 1$
and Fermi wavenumber $k_F=2\pi^2\rho_0^2 n \gg 1$ [or $n\gg
(2\pi^3\rho_0^2 a_0)^{-1}$]. \citet{skud93} use the random-phase
approximation to find a numerical fit for the correlation energy in
this regime (see also Ref.~\citep{stein98}):
\be \varepsilon_{\rm corr} =
-\frac{e^2}{\rho_0}\,[0.595 (t/b)^{1/8} (1-1.009t^{1/8})] \,.
\label{correq}
\ee
In the absence of an ``exact'' correlation energy density we employ
this strong-field-limit expression. Fortunately, because we are
concerned mostly with finding energy changes between different states
of atoms and molecules, the correlation energy term does not have to
be exact. The presence or the form of the correlation term has a
modest effect on the atomic and molecular energies calculated but has
very little effect on the energy difference between them (see Appendix
B for more details on various forms of the correlation energy and
comparisons).

Variation of the total energy with respect to the total electron
density, $\delta E[n]/\delta n=0$, leads to the Kohn-Sham equations:
\be 
\left[ -\frac{\hbar^2}{2m_e} \nabla^2 + V_{\rm eff}(\mathbf{r}) \right]
\Psi_{m\nu}(\mathbf{r}) = \varepsilon_{m\nu} \Psi_{m\nu}(\mathbf{r}) \,, 
\ee
where
\be
V_{\rm eff}(\mathbf{r}) =
-\!\!\sum_{j=1}^N \frac{Ze^2}{|\mathbf{r}-\mathbf{z}_j|}
+ e^2 \int d\mathbf{r}' \, \frac{n(\mathbf{r}')}{|\mathbf{r} - \mathbf{r}'|}
+ \mu_{\rm exc}(n),
\ee
with
\be
\mu_{\rm exc}(n) = \frac{\partial (n \varepsilon_{\rm exc})}{\partial n} \,.
\ee
Averaging the Kohn-Sham equations over the transverse wave function yields
a set of one-dimensional equations:
\ba 
\left( -\frac{\hbar^2}{2m_e}\frac{d^2}{dz^2}
- \!\!\sum_{j=1}^N Ze^2 \int d\mathbf{r}_\perp \,
\frac{|W_m|^2(\rho)}{|\mathbf{r}-\mathbf{z}_j|}
+ e^2 \int \!\! \int d\mathbf{r}_\perp\,d\mathbf{r}' \,
\frac{|W_m|^2(\rho)\, n(\mathbf{r}')}{|\mathbf{r} - \mathbf{r}'|}
\right. \nonumber\\
\left. + \int d\mathbf{r}_\perp \, |W_m|^2(\rho)\, \mu_{\rm exc}(n) \right)
f_{m\nu}(z) = \varepsilon_{m\nu} f_{m\nu}(z) \,.
\label{kohneq}
\ea
These equations are solved self-consistently to find the eigenvalue
$\varepsilon_{m\nu}$ and the longitudinal wave function
$f_{m\nu}(z)$ for each orbital occupied by the $ZN$ electrons.
Once these are known, the total energy of the system 
can be calculated using 
\be 
E[n] = \sum_{m\nu} \varepsilon_{m\nu} -
\frac{e^2}{2} \int \!\! \int d\mathbf{r} d\mathbf{r}' \,
\frac{n(\mathbf{r}) n(\mathbf{r}')}{|\mathbf{r} - \mathbf{r}'|} + \int
d\mathbf{r} \, n(\mathbf{r}) [\varepsilon_{\rm exc}(n)-\mu_{\rm exc}(n)] +
\sum_{j=1}^{N-1} (N-j)\frac{Z^2e^2}{ja} \,.
\label{energyeq}
\ee
Details of our method used in computing the various integrals and solving
the above equations are given in Appendix A\@.

Note that for a given system, the occupations of electrons in
different $(m\nu)$ orbitals are not known {\it a priori}, and must be
determined as part of the procedure of finding the minimum energy
state of the system.  In our calculation, we first guess
$n_0,n_1,n_2,\ldots$, the number of electrons in the
$\nu=0,~1,~2,\ldots$ orbitals, respectively (e.g., the electrons in
the $\nu=0$ orbitals have $m=0,1,2,\ldots, n_0-1$).  Note that
$n_0+n_1+n_2+\cdots=NZ$. We find the energy of the system for this
particular set of electron occupations. We then vary the electron
occupations and repeat the calculation until the true minimum energy
state is found. Obviously, in the case of molecules, we must vary the
ion spacing $a$ to determine the equilibrium separation and the the
ground-state energy of the molecule.  Graphical examples of how the
ground state is chosen are given in the next section.

\section{Results}

In this section we present our results for the parallel configuration
of H$_N$ (up to $N=10$), He$_N$ (up to $N=8$), C$_N$ (up to $N=5$),
and Fe$_N$ (up to $N=3$) at various magnetic field strengths between
$B = 10^{12}$~G and $2\times10^{15}$~G\@. For each molecule (or atom),
data is given in tabular form on the molecule's ground-state energy,
the equilibrium separation of the ions in the molecule, and its
orbital structure (electron occupation numbers
$n_0,n_1,n_2,\ldots$). In some cases the first-excited-state energies
are given as well, when the ground-state and first-excited-state
energies are similar in value. We also provide the ground-state
energies for selected ionization states of C and Fe atoms; among other
uses, these quantities are needed for determining the ion emission
from a condensed neutron star surface \citep{medin06a}. All of the
energies presented in this section are calculated to better than
$0.1\%$ numerical accuracy (see Appendix A).

For each of the molecules and ions presented in this section we
provide numerical scaling relations for the ground-state energy as a
function of magnetic field, in the form of a scaling exponent $\beta$
with $E_N \propto B_{12}^\beta$. We have provided this information to
give readers easy access to energy values for fields in between those
listed in the tables. The ground-state energy is generally not
well fit by a constant $\beta$ over the entire magnetic field range
covered by this work, so we have provided $\beta$ values over several
different magnetic field ranges. Note that the theoretical value
$\beta=2/5$ (see Sec.~II) is approached only in certain asymptotic
limits.

We discuss here briefly a few trends in the data: All of the molecules
listed in the following tables are bound. The Fe$_2$ and Fe$_3$
molecules at $B_{12}=5$ are not bound, so we have not listed them
here, but we have listed the Fe atom at this field strength for
comparison with other works. All of the bound molecules listed below
have ground-state energies per atom that decrease monotonically with
increasing $N$, with the exception of H$_N$ at $B_{12}=1$, which has a
slight upward glitch in energy at H$_4$ (see
Table~\ref{Htable}). Additionally, these energies approach asymptotic
values for large $N$ --- the molecule essentially becomes
one-dimensional condensed matter \citep{medin06a}. The equilibrium ion
separations also approach asymptotic values for large $N$, but there
is no strong trend in the direction of approach: sometimes the
equilibrium ion separations increase with increasing $N$, sometimes
they decrease, and sometimes they oscillate back and forth.

In general, we find that for a given molecule (e.g., Fe$_3$), the
number of electrons in $\nu>0$ states decreases as the magnetic field
increases. This is because the characteristic transverse size $\rho_0
\propto B^{-1/2}$ decreases, so the electrons prefer to stay in the
$\nu=0$ states. For a given field strength, as the number of electrons
in the system $NN_e$ increases (e.g., from Fe$_2$ to Fe$_3$), more
electrons start to occupy the $\nu>0$ states since the average
electron-nucleus separation $\rho_m \propto (2m+1)^{1/2} B^{-1/2}$
becomes too large for large $m$. For large enough $N$ the value of
$n_0$, the number of electrons in $\nu=0$ states, levels off,
approaching its infinite chain value (see
Ref.~\citep{medin06a}). Similar trends happen with $n_1$, $n_2$, etc.,
though much more slowly.

There are two ways that we have checked the validity of our results by
comparison with other works. First, we have repeated several of our
atomic and molecular calculations using the correlation energy
expression empirically determined by \citet{jones85}:
\be
\varepsilon_{\rm corr} = -\frac{e^2}{\rho_0}(0.0096 \ln \rho_0^3 n + 0.122) \,.
\label{Joneseq}
\ee
The results we then obtain for the atomic ground-state energies
agree with those of \citet{jones85,jones86}. For example, for Fe at
$B_{12}=5$ we find an atomic energy of $-108.05$~eV and Jones gives an
energy of $-108.18$~eV\@. The molecular ground-state energies per atom
are of course not the same as those for the infinite chain from
Jones's work, but they are comparable, particularly for the large
molecules. For example, we find for He$_8$ at $B_{12}=5$ that the
energy per atom is $-1242$~eV and Jones finds for He$_\infty$ that the
energy per cell is $-1260$~eV\@. (See Appendix B for a brief
discussion of why in our calculations we chose to use the
Skudlarski-Vignale correlation energy expression over that of Jones.)

Second, we have compared our hydrogen, helium, and carbon molecule
results to those of Refs.~\citep{demeur94,lai92}. Because these works
use the Hartree-Fock method, we cannot compare absolute ground-state
energies with theirs, but we can compare energy differences. We find
fair agreement, though the Hartree-Fock results are consistently
smaller. Some of these comparisons are presented in the following
subsections.

\subsection{Hydrogen}

\begin{figure}
\includegraphics[width=6.5in]{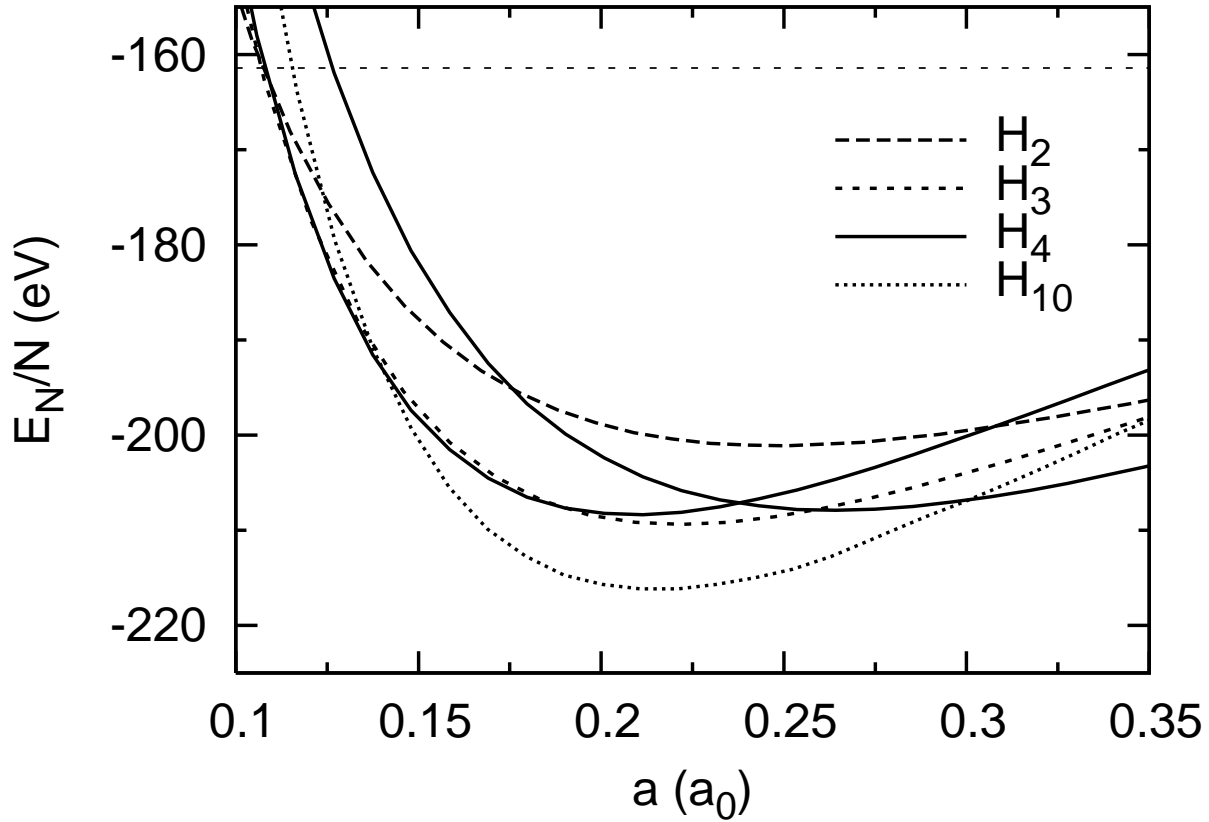}
\caption{Molecular energy per atom versus ion separation for various
hydrogen molecules at $B_{12} = 1$. The energy of the H atom is shown
as a horizontal line at $-161.4$~eV\@. The two lowest-energy
configurations of H$_4$ have nearly the same minimum energy, so the
curves for both configurations are shown here.}
\label{HMolfig}
\end{figure}

Our numerical results for H are given in Table~\ref{Htable} and
Table~\ref{Htable2}. Note that at $B_{12}=1$, H$_4$ is less bound than
H$_3$, and thus $E=E_N/N$ is not a necessarily a monotonically
decreasing function of $N$ at this field strength. For the H$_4$
molecule, two configurations, $(n_0,n_1)=(4,0)$ and $(3,1)$, have very
similar equilibrium energies (see Fig.~\ref{HMolfig}), although the
equilibrium ion separations are different. The real ground state may
therefore be a ``mixture'' of the two configurations; such a state
would presumably give a lower ground-state energy for H$_4$, and make
the energy trend monotonic.

Hartree-Fock results for H molecules are given in \citep{lai92}.
For H$_2$, H$_3$, and H$_4$, the energies (per atom) are, respectively: 
$-184.3$,~$-188.7$,~$-185.0$~eV at $B_{12}=1$;
$-383.9$,~$-418.8$,~$-432.9$~eV at $B_{12}=10$; and 
$-729.3$,~$-847.4$,~$-915.0$~eV at $B_{12}=100$.
Thus, our density-functional-theory calculation
tends to overestimate the energy $|E|$ by about $10\%$. Note that
the Hartree-Fock results also reveal a non-monotonic behavior
of $E$ at $N=4$ for $B_{12}=1$, in agreement with our density-functional
result. \citet{demeur94} calculated the energies of
H$_2$--H$_5$ at $B_{12}=2.35$; their results exhibit similar trends.

\begin{table}
\caption{Ground-state energies, ion separations, and electron
configurations of hydrogen molecules, over a range of magnetic field
strengths. In some cases the first-excited-state energies are also
listed. Energies are given in units of eV, separations in units of
$a_0$ (the Bohr radius). For molecules (H$_N$) the energy per atom is
given, $E = E_N/N$. All of the H and H$_2$ molecules listed here have
electrons only in the $\nu=0$ states. For the H$_3$ and larger
molecules here, however, the molecular structure is more complicated,
and is designated by the notation $(n_0,n_1,\ldots)$, where $n_0$
is the number of electrons in the $\nu=0$ orbitals, $n_1$ is the
number of electrons in the $\nu=1$ orbitals, etc.}
\label{Htable}
\centering
\begin{tabular}{c | r@{}l | r@{}l l | r@{}l l c | r@{}l l c | r@{}l l c}
\hline\hline
 & \multicolumn{2}{| c |}{H} & \multicolumn{3}{| c |}{H$_2$} & \multicolumn{4}{| c |}{H$_3$} & \multicolumn{4}{| c |}{H$_4$} & \multicolumn{4}{| c}{H$_5$} \\
$B_{12}$ & \multicolumn{2}{| c}{$E$} & \multicolumn{2}{| c}{$E$} & \multicolumn{1}{c}{$a$} & \multicolumn{2}{| c}{$E$} & \multicolumn{1}{c}{$a$} & $(n_0,n_1)$ & \multicolumn{2}{| c}{$E$} & \multicolumn{1}{c}{$a$} & $(n_0,n_1)$ & \multicolumn{2}{| c}{$E$} & \multicolumn{1}{c}{$a$} & $(n_0,n_1)$ \\
\hline
1 & -161&.4 & -201&.1 & 0.25 & -209&.4 & 0.22 & (3,0) & -208&.4 & 0.21 & (4,0) & -213&.8 & 0.23 & (4,1) \\
 & && && & -191&.1 & 0.34 & (2,1) & -207&.9 & 0.26 & (3,1) & -203&.1 & 0.200 & (5,0) \\
\hline
10 & -309&.5 & -425&.8 & 0.125 & -469&.0 & 0.106 & (3,0) & -488&.1 & 0.096 & (4,0) & -493&.5 & 0.090 & (5,0) \\
 & && && & && & & && & & -478&.9 & 0.112 & (4,1) \\
\hline
100 & -540&.3 & -829&.5 & 0.071 & -961&.2 & 0.057 & (3,0) & -1044&.5 & 0.049 & (4,0) & -1095&.5 & 0.044 & (5,0) \\
\hline
1000 & -869&.6 & -1540&.5 & 0.044 & -1818&.0 & 0.033 & (3,0) & -2049& & 0.028 & (4,0) & -2222& & 0.024 & (5,0) \\
\hline\hline
\end{tabular}
\\[10pt]
\begin{tabular}{c | r@{}l l c | r@{}l l c | r@{}l l c}
\hline\hline
 & \multicolumn{4}{| c |}{H$_6$} & \multicolumn{4}{| c |}{H$_8$} & \multicolumn{4}{| c}{H$_{10}$} \\
$B_{12}$ & \multicolumn{2}{| c}{$E$} & \multicolumn{1}{c}{$a$} & $(n_0,n_1)$ & \multicolumn{2}{| c}{$E$} & \multicolumn{1}{c}{$a$} & $(n_0,n_1,n_2)$ & \multicolumn{2}{| c}{$E$} & \multicolumn{1}{c}{$a$} & $(n_0,n_1,n_2)$ \\
\hline
1 & -214&.1 & 0.23 & (4,2) & -215&.8 & 0.23 & (5,2,1) & -216&.2 & 0.22 & (6,3,1) \\
 & -213&.4 & 0.21 & (5,1) & -215&.3 & 0.25 & (4,3,1) & -216&.0 & 0.23 & (5,3,2) \\
\hline
10 & -496&.5 & 0.101 & (5,1) & -507&.1 & 0.095 & (8,2,0) & -509&.3 & 0.091 & (7,3,0) \\
 & -490&.8 & 0.86 & (6,0) & -504&.1 & 0.089 & (7,1,0) & -506&.8 & 0.087 & (8,2,0) \\
\hline
100 & -1125&.0 & 0.041 & (6,0) & -1143&.0 & 0.038 & (8,0,0) & -1169&.5 & 0.038 & (9,1,0) \\
 & && & & -1139&.5 & 0.043 & (7,1,0) & -1164&.0 & 0.042 & (8,2,0) \\
\hline
1000 & -2351& & 0.22 & (6,0) & -2518& & 0.0190 & (8,0,0) & -2600& & 0.0170 & (10,0,0) \\
 & && & & && & & -2542& & 0.0200 & (9,1,0) \\
\hline\hline
\end{tabular}
\end{table}

\begin{table}
\caption{Fit of the ground-state energies of hydrogen molecules to the
scaling relation $E \propto B_{12}^\beta$. The scaling exponent
$\beta$ is fit for each molecule H$_N$ over three magnetic field
ranges: $B_{12}=1-10$, $10-100$, and $100-1000$.}
\label{Htable2}
\centering
\begin{tabular}{c | c c c c c c c c}
\hline\hline
 & \multicolumn{8}{| c}{$\beta$} \\
$B_{12}$ & H & H$_2$ & H$_3$ & H$_4$ & H$_5$ & H$_6$ & H$_8$ & H$_{10}$ \\
\hline
1-10 & 0.283 & 0.326 & 0.350 & 0.370 & 0.363 & 0.365 & 0.371 & 0.372 \\
\hline
10-100 & 0.242 & 0.290 & 0.312 & 0.330 & 0.346 & 0.355 & 0.353 & 0.361 \\
\hline
100-1000 & 0.207 & 0.269 & 0.277 & 0.293 & 0.307 & 0.320 & 0.343 & 0.347 \\
\hline\hline
\end{tabular}
\end{table}

\subsection{Helium}

Our numerical results for He are given in
Table~\ref{Hetable} and Table~\ref{Hetable2}.

The energies (per atom) of He and He$_2$ based on Hartree-Fock
calculations \citep{lai01} are, respectively: $-575.5$,~$-601.2$~eV at
$B_{12}=1$; $-1178$,~$-1364$~eV at $B_{12}=10$; $-2193$,~$-2799$~eV at
$B_{12}=100$; and $-3742$,~$-5021$~eV at $B_{12}=1000$.  At
$B_{12}=2.35$, \citet{demeur94} find that the energies (per atom) of
He, He$_2$, He$_3$, and He$_4$ are, respectively:
$-753.4$,~$-812.6$,~$-796.1$,~$-805.1$~eV\@. Using our scaling
relations, we find for that same field that the energies of He,
He$_2$, He$_3$, and He$_5$ (we do not have an He$_4$ result) are:
$-791$,~$-871$,~$-889$,~$-901$~eV\@.  Thus, our density-functional
theory calculation tends to overestimate the energy $|E|$ by about
$10\%$.

\begin{table}
\caption{Ground-state energies, ion separations, and electron
configurations of helium molecules, over a range of magnetic field
strengths. In some cases the first-excited-state energies are also
listed. Energies are given in units of eV, separations in units of
$a_0$ (the Bohr radius). For molecules (He$_N$) the energy per atom is
given, $E = E_N/N$. All of the He and He$_2$ molecules listed here
have electrons only in the $\nu=0$ states. For the He$_3$ and larger
molecules here, however, the molecular structure is more complicated,
and is designated by the notation $(n_0,n_1,\ldots)$, where $n_0$
is the number of electrons in the $\nu=0$ orbitals, $n_1$ is the
number of electrons in the $\nu=1$ orbitals, etc.}
\label{Hetable}
\centering
\begin{tabular}{c | r@{}l | r@{}l l | r@{}l l c | r@{}l l c | r@{}l l c}
\hline\hline
 & \multicolumn{2}{| c |}{He} & \multicolumn{3}{| c |}{He$_2$} & \multicolumn{4}{| c |}{He$_3$} & \multicolumn{4}{| c |}{He$_5$} & \multicolumn{4}{| c}{He$_8$} \\
$B_{12}$ & \multicolumn{2}{| c}{$E$} & \multicolumn{2}{| c}{$E$} & \multicolumn{1}{c}{$a$} & \multicolumn{2}{| c}{$E$} & \multicolumn{1}{c}{$a$} & $(n_0,n_1)$ & \multicolumn{2}{| c}{$E$} & \multicolumn{1}{c}{$a$} & $(n_0,n_1,n_2)$ & \multicolumn{2}{| c}{$E$} & \multicolumn{1}{c}{$a$} & $(n_0,n_1,n_2,n_3)$ \\
\hline
1 & -603&.5 & -641&.2 & 0.25 & -647&.3 & 0.28 & (5,1) & -653&.1 & 0.29 & (6,3,1) & -656&.7 & 0.28 & (7,5,3,1) \\
 & && && & -633&.0 & 0.32 & (4,2) & -649&.4 & 0.28 & (7,2,1) & -656&.5 & 0.27 & (8,5,2,1) \\
\hline
10 & -1252&.0 & -1462&.0 & 0.115 & -1520&.0 & 0.105 & (6,0) & -1553&.5 & 0.110 & (8,2,0) & -1574&.5 & 0.110 & (10,5,1,0) \\
 & && && & -1462&.0 & 0.125 & (5,1) & -1547&.5 & 0.105 & (9,1,0) & -1574&.0 & 0.105 & (11,4,1,0) \\
\hline
100 & -2385& & -3039& & 0.060 & -3370& & 0.050 & (6,0) & -3573& & 0.044 & (10,0,0) & -3694& & 0.045 & (13,3,0,0) \\
 & && && & -3140& & 0.054 & (5,1) & -3543& & 0.049 & (9,1,0) & -3690& & 0.043 & (14,2,0,0) \\
\hline
1000 & -4222& & -5787& & 0.036 & -6803& & 0.028 & (6,0) & -7887& & 0.022 & (10,0,0) & -8406& & 0.0200 & (15,1,0,0) \\
 & && && & && & & && & & -8357& & 0.0180 & (16,0,0,0) \\
\hline\hline
\end{tabular}
\end{table}

\begin{table}
\caption{Fit of the ground-state energies of helium molecules to the
scaling relation $E \propto B_{12}^\beta$. The scaling exponent
$\beta$ is fit for each molecule He$_N$ over three magnetic field
ranges: $B_{12}=1-10$, $10-100$, and $100-1000$.}
\label{Hetable2}
\centering
\begin{tabular}{c | c c c c c}
\hline\hline
 & \multicolumn{5}{| c}{$\beta$} \\
$B_{12}$ & He & He$_2$ & He$_3$ & He$_5$ & He$_8$ \\
\hline
1-10 & 0.317 & 0.358 & 0.371 & 0.376 & 0.380 \\
\hline
10-100 & 0.280 & 0.318 & 0.346 & 0.362 & 0.370 \\
\hline
100-1000 & 0.248 & 0.280 & 0.305 & 0.344 & 0.357 \\
\hline\hline
\end{tabular}
\end{table}

\subsection{Carbon}

Our numerical results for C are given in
Table~\ref{Ctable}, Table~\ref{Ctable2}, and Table~\ref{Ctable3}.

The only previous result of C molecules is that by \citet{demeur94},
who calculated C$_2$ only at $B_{12}=2.35$.  At this field strength,
our calculation shows that C$_2$ is bound relative to C atom
($E=-5994$,~$-6017$~eV for C, C$_2$), whereas Demeur {\it et al.} find no
binding ($E=-5770$,~$-5749$~eV for C, C$_2$).  Thus our result differs
qualitatively from \citep{demeur94}. We also disagree on the
ground-state occupation at this field strength: we find
$(n_0,n_1)=(9,3)$ while Demeur {\it et al.}\ find $(n_0,n_1)=(7,5)$. We
suggest that if Demeur {\it et al.}\ used the occupation $(n_0,n_1)=(9,3)$
they would obtain a lower-energy for C$_2$, though whether C$_2$ would
then be bound remains uncertain. Since the numerical accuracy of our
computation is $0.1\%$ of the total energy (thus, about $6$~eV for
$B_{12}=2.35$), our results for $B_{12}\alt$ a few should be treated
with caution.

\begin{figure}
\includegraphics[width=6.5in]{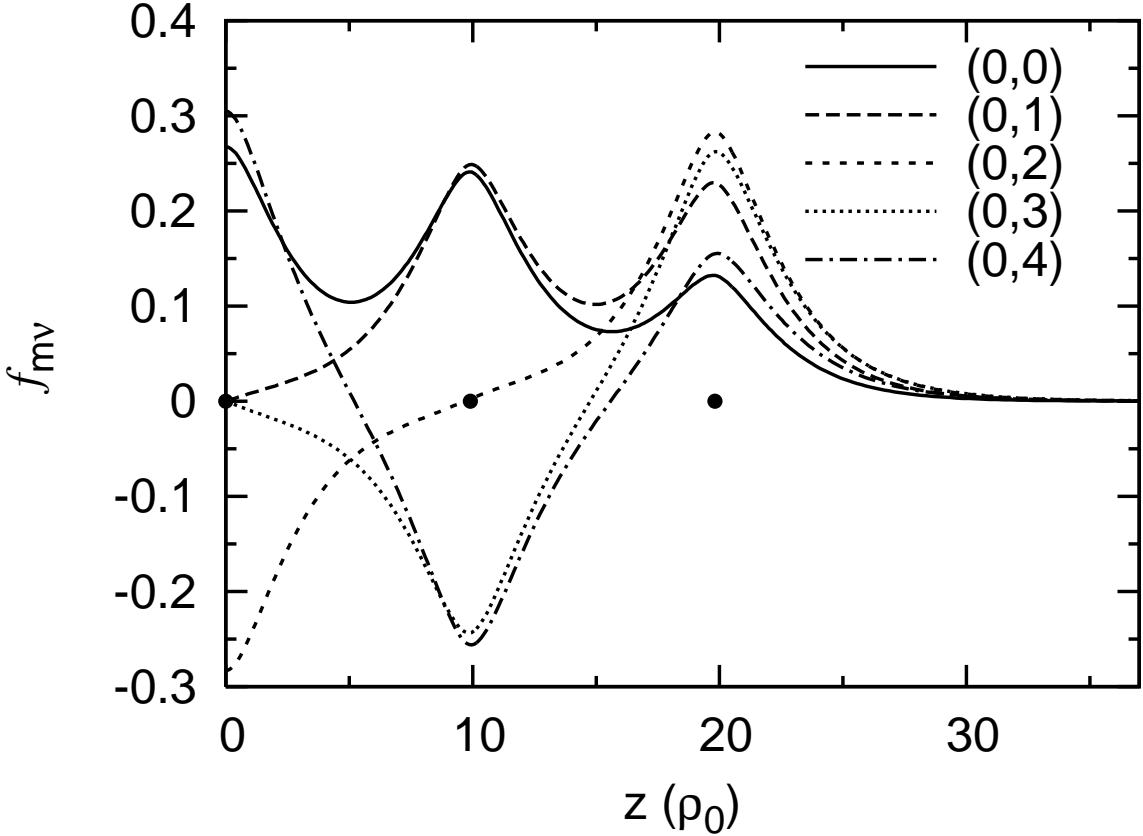}
\caption{Longitudinal wave functions for selected electron orbitals of
C$_5$ at $B_{12} = 1$, at the equilibrium ion separation. Different
orbitals are labeled by $(m,\nu)$. Only the $z \ge 0$ region is
shown. Wave Functions with even $\nu$ are symmetric about $z=0$, and
those with odd $\nu$ are antisymmetric about $z=0$. The filled circles
denote the ion locations.}
\label{wfCfig}
\end{figure}

Figure~\ref{wfCfig} gives some examples of the longitudinal electron
wave functions. One wave function of each node type in the molecule
($\nu=0$ to $4$) is represented. Note that on the atomic scale each
wave function is nodeless in nature; that is, there are no nodes at the
ions, only in between ions. The exception to this is at the central
ion, where due to symmetry considerations the antisymmetric
wave functions must have nodes. [The nodes for $(m,\nu)=(0,2)$ are
near, but not at, the ions $j=2$ and $j=4$. This is incidental.] This
is not surprising when one considers that all of the electrons in
atomic carbon at this field strength are nodeless. The entire
molecular wave function can be thought of as a string of atomic
wave functions, one around each ion, each modified by some phase factor
to give the overall nodal nature of the wave function. Indeed, for
atoms at field strengths that are low enough to allow $\nu>0$ states,
we find that their corresponding molecules have electron wave functions
with nodes at the ions. Atomic Fe at $B_{12}=10$, for example, has an
electron wave function with one node at the ion, and Fe$_2$ at
$B_{12}=10$ has an electron wave function with a node at each ion.

\begin{table}
\caption{Ground-state energies, ion separations, and electron
configurations of carbon molecules, over a range of magnetic field
strengths. In some cases the first-excited-state energies are also
listed. Energies are given in units of eV, separations in units of
$a_0$ (the Bohr radius). For molecules (C$_N$) the energy per atom is
given, $E = E_N/N$. All of the C atoms listed here have electrons only
in the $\nu=0$ orbitals. For the C$_2$ and larger molecules here,
however, the molecular structure is more complicated, and is
designated by the notation $(n_0,n_1,\ldots)$, where $n_0$ is the
number of electrons in the $\nu=0$ orbitals, $n_1$ is the number of
electrons in the $\nu=1$ orbitals, etc.}
\label{Ctable}
\centering
\begin{tabular}{c | r | r l c | r l c | r l c | r l c}
\hline\hline
 & \multicolumn{1}{| c |}{C} & \multicolumn{3}{| c |}{C$_2$} & \multicolumn{3}{| c |}{C$_3$} & \multicolumn{3}{| c |}{C$_4$} & \multicolumn{3}{| c}{C$_5$} \\
$B_{12}$ & \multicolumn{1}{| c}{$E$} & \multicolumn{1}{| c}{$E$} & \multicolumn{1}{c}{$a$} & $(n_0,n_1)$ & \multicolumn{1}{| c}{$E$} & \multicolumn{1}{c}{$a$} & $(n_0,n_1,n_2)$ & \multicolumn{1}{| c}{$E$} & \multicolumn{1}{c}{$a$} & $(n_0,n_1,n_2,n_3)$ & \multicolumn{1}{| c}{$E$} & \multicolumn{1}{c}{$a$} & $(n_0,n_1,n_2,n_3,n_4)$ \\
\hline
1 & -4341 & -4351 & 0.53 & (8,4) & -4356 & 0.52 & (9,6,3) & -4356 & 0.52 & (10,7,4,3) & -4358 & 0.48 & (11,8,6,3,2) \\
 & & -4349 & 0.46 & (9,3) & -4354 & 0.50 & (10,5,3) & -4354 & 0.56 & (9,7,5,3) & -4357 & 0.47 & (12,8,5,3,2) \\
\hline
10 & -10075 & -10215 & 0.150 & (11,1) & -10255 & 0.175 & (13,4,1) & -10255 & 0.180 & (15,6,2,1) & -10275 & 0.150 & (18,8,3,1) \\
 & & -10200 & 0.180 & (10,2) & -10240 & 0.185 & (14,3,1) & -10250 & 0.185 & (14,7,2,1) & -10270 & 0.155 & (17,9,3,1) \\
\hline
100 & -21360 & -23550 & 0.054 & (12,0) & -24060 & 0.055 & (17,1,0) & -24350 & 0.054 & (21,3,0,0) & -24470 & 0.057 & (23,6,1,0,0) \\
 & & & & & -23960 & 0.058 & (16,2,0) & -24300 & 0.056 & (20,4,0,0) & -24460 & 0.056 & (24,5,1,0,0) \\
\hline
1000 & -41330 & -50760 & 0.027 & (12,0) & -54870 & 0.024 & (18,0,0) & -56500 & 0.024 & (23,1,0,0) & -57640 & 0.022 & (28,2,0,0,0) \\
 & & & & & & & & -56190 & 0.022 & (24,0,0,0) & -57520 & 0.023 & (27,3,0,0,0) \\
\hline\hline
\end{tabular}
\end{table}

\begin{table}
\caption{Ground-state energies of ionized carbon atoms over a range of
magnetic field strengths. Energies are given in units of eV\@. For
these field strengths, the electron configuration of C atoms is such
that all of their electrons lie in the $\nu=0$ orbitals; therefore the
ionized atoms have all electrons in the $\nu=0$ orbitals as well. The
ionization state is designated by the notation, ``C$^{n+}$,'' where
$n$ is the number of electrons that have been removed from the
atom. The entry ``C$^{5+}$,'' for example, is a carbon nucleus plus
one electron.}
\label{Ctable2}
\centering
\begin{tabular}{c | r | r | r | r | r | r@{}l}
\hline\hline
$B_{12}$ & \multicolumn{1}{| c}{C} & \multicolumn{1}{| c}{C$^{+}$} & \multicolumn{1}{| c}{C$^{2+}$} & \multicolumn{1}{| c}{C$^{3+}$} & \multicolumn{1}{| c}{C$^{4+}$} & \multicolumn{2}{| c}{C$^{5+}$} \\
\hline
1 & -4341 & -4167 & -3868 & -3411 & -2739 & -1738&.0 \\
\hline
10 & -10075 & -9644 & -8917 & -7814 & -6213 & -3877& \\
\hline
100 & -21360 & -20370 & -18730 & -16300 & -12815 & -7851& \\
\hline
1000 & -41330 & -39210 & -35830 & -30920 & -24040 & -14425& \\
\hline\hline
\end{tabular}
\end{table}

\begin{table}
\caption{Fit of the ground-state energies of neutral and ionized
carbon atoms and carbon molecules to the scaling relation $E \propto
B_{12}^\beta$. The scaling exponent $\beta$ is fit over three magnetic
field ranges: $B_{12}=1-10$, $10-100$, and $100-1000$.}
\label{Ctable3}
\centering
\begin{tabular}{c | c c c | c | c c c c}
\hline\hline
 & \multicolumn{7}{c}{$\beta$} & \\
$B_{12}$ & C$^{5+}$ & C$^{4+}$ & C$^{+}$ & C & C$_2$ & C$_3$ & C$_4$ & C$_5$ \\
\hline
1-10 & 0.348 & 0.356 & 0.364 & 0.366 & 0.371 & 0.372 & 0.372 & 0.372 \\
\hline
10-100 & 0.306 & 0.314 & 0.325 & 0.326 & 0.363 & 0.370 & 0.376 & 0.377 \\
\hline
100-1000 & 0.264 & 0.273 & 0.284 & 0.287 & 0.334 & 0.358 & 0.366 & 0.372 \\
\hline\hline
\end{tabular}
\end{table}

\subsection{Iron}

\begin{figure}
\includegraphics[width=6.5in]{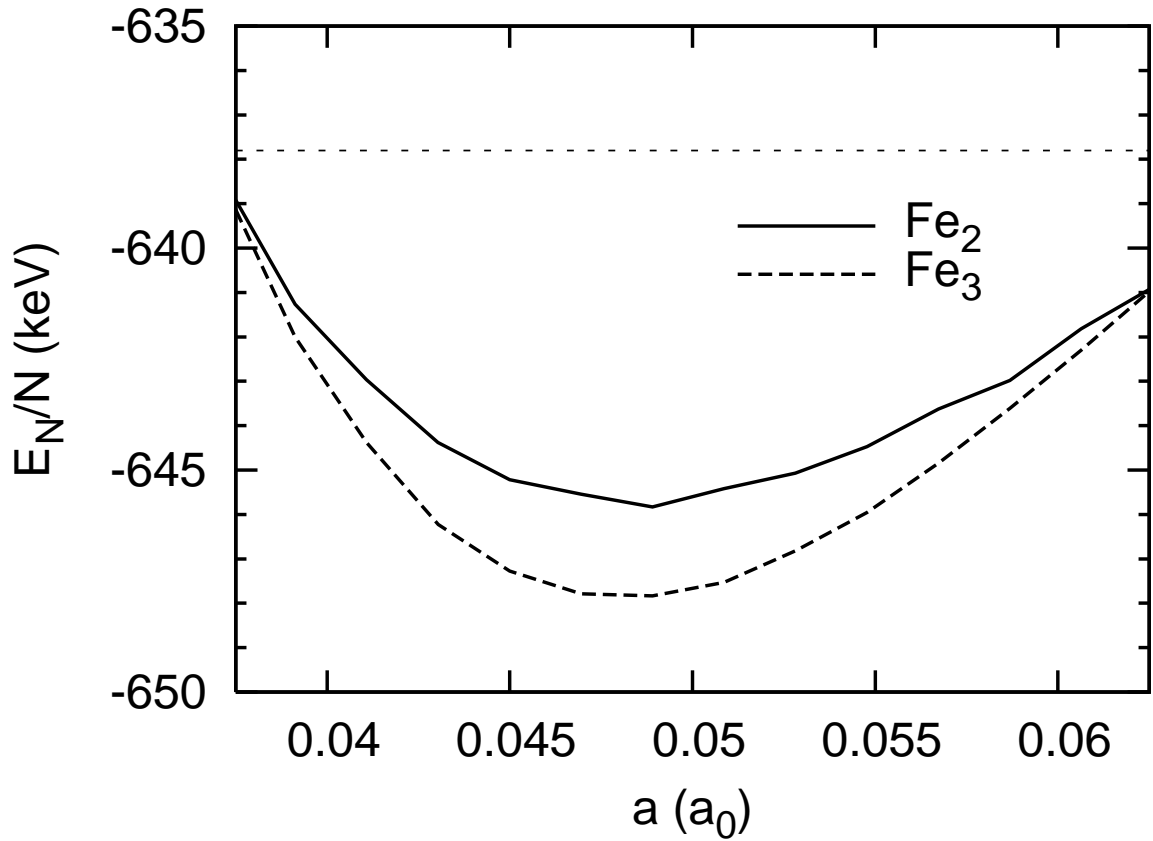}
\caption{Molecular energy per atom versus ion separation for Fe$_2$
and Fe$_3$ molecules at $B_{12} = 500$. The energy of the Fe atom is
shown as a horizontal line at $-637.8$~keV\@.}
\label{FeMol500fig}
\end{figure}

\begin{figure}
\includegraphics[width=6.5in]{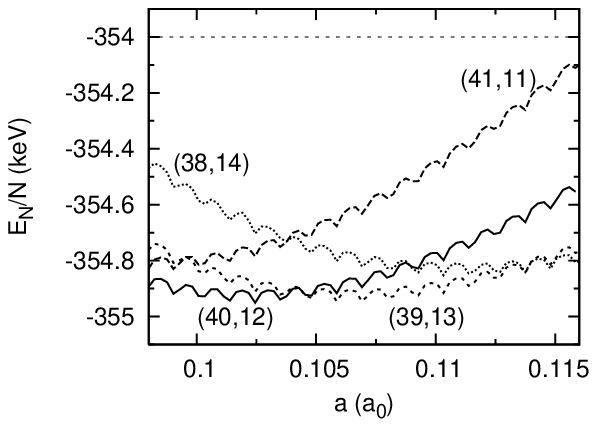}
\caption{Molecular energy per atom versus ion separation for various
configurations of electrons in the Fe$_2$ molecule at $B_{12} =
100$. The configurations are labeled using the notation
$(n_0,n_1)$, where $n_0$ is the number of electrons with $\nu=0$
and $n_1$ is the number with $\nu=1$. The energy of the Fe atom is
shown as a horizontal line at $-354.0$~keV\@. The states ``$(40,12)$''
and ``$(39,13)$'' have the lowest equilibrium energies of all possible
configurations and within the numerical accuracy of our calculation
have the same equilibrium energies. The wavy structure of the curves
gives an indication of the numerical accuracy of our code. Note that
states with electrons in the $\nu=2$ orbitals [for example,
$(39,12,1)$] have energies higher than the atomic energy and are
therefore unbound.}
\label{minfig}
\end{figure}

Our numerical results for Fe are given in Table~\ref{Fetable},
Table~\ref{Fetable2}, and Table~\ref{Fetable3}. The energy curves for
$B_{12}=500$ are shown in Fig.~\ref{FeMol500fig}, and some results for
$B_{12}=100$ are shown in Fig.~\ref{minfig}.

There is no previous quantitative calculation of Fe molecules in
strong magnetic fields that we are aware of. The most relevant work is
that of \citet{abrahams91}, who use a Thomas-Fermi type model to
calculate Fe and Fe$_2$ energies for magnetic fields up to
$B_{12}=30$. Unfortunately, a comparison of our results with those of
this work is not very useful, as Thomas-Fermi models are known to give
inaccurate energies and in particular large overestimates of binding
and cohesive energies. As an example, from Ref.~\citep{abrahams91} the
energy difference between Fe and Fe$_2$ at $B_{12}=30$ is $1.7$~keV,
which is twice as large as our result at $B_{12}=100$.

In Table~\ref{Fetable} we have not provided results for the Fe$_2$ and
Fe$_3$ molecules at $B_{12}=5$, as these molecules are not bound
relative to the Fe atom. We have not provided results for the Fe$_3$
molecule at $B_{12}=10$ because the energy difference (per atom)
between Fe$_3$ and the Fe atom at this field strength is smaller than
the error in our calculation, $0.1\%$ of $|E|$ or $140$~eV\@. The
energy difference (per atom) between the Fe$_2$ molecule and the Fe
atom at $B_{12}=10$ is also smaller than the error in our calculation
(indeed, the difference should be less than that between Fe$_3$ and Fe
at this field strength), but we have redone the calculation using more
grid and integration points such that the energy values reported here
for these two molecules are accurate numerically to $0.01\%$ (see
Appendix A). At this accuracy, our results indicate that Fe$_2$ is
bound over Fe at $B_{12}=10$ with a energy difference per atom of
$30$~eV\@.

Figure~\ref{minfig} illustrates how the ground-state electron
configuration is found for each molecule. The configuration with the
lowest equilibrium energy is chosen as the ground-state
configuration. In the case depicted in Fig.~\ref{minfig}, Fe$_2$ at
$B_{12}=100$, there are actually two such configurations. Within the
error of our calculation, we cannot say which one represents the
ground state. Note that the systematic error seen in the minimization
curves of the various Fe$_2$ configurations is much smaller than our
target $0.1\%$ error for the total energy (the sinusoidal error in the
figure has an amplitude of $\approx 30$~eV, or around $0.01\%$ of the
total energy).

\begin{table}
\caption{Ground-state energies, ion separations, and electron
configurations of iron molecules, over a range of magnetic field
strengths. In some cases the first-excited-state energies are also
listed. Energies are given in units of keV, separations in units of
$a_0$ (the Bohr radius). For molecules (Fe$_N$) the energy per atom is
given, $E = E_N/N$. The electron configuration is designated by the
notation $(n_0,n_1,\ldots)$, where $n_0$ is the number of
electrons in the $\nu=0$ orbitals, $n_1$ is the number of electrons in
the $\nu=1$ orbitals, etc. Note that no information is listed for the
Fe$_2$ and Fe$_3$ molecules at $B_{12}=5$, as we have found that these
molecules are not bound at this field strength. Also note that there
are two lowest-energy states for Fe$_2$ at $B_{12}=100$; within the
error of our calculation, the two states have the same minimum
eneriges.}
\label{Fetable}
\centering
\begin{tabular}{c | r@{}l c | r@{}l l c | r@{}l l c}
\hline\hline
 & \multicolumn{3}{| c |}{Fe} & \multicolumn{4}{| c |}{Fe$_2$} & \multicolumn{4}{| c}{Fe$_3$} \\
$B_{12}$ & \multicolumn{2}{| c}{$E$} & $(n_0,n_1)$ & \multicolumn{2}{| c}{$E$} & \multicolumn{1}{c}{$a$} & $(n_0,n_1)$ & \multicolumn{2}{| c}{$E$} & \multicolumn{1}{c}{$a$} & $(n_0,n_1,n_2)$ \\
\hline
5 & -107&.20 & (24,2) & \multicolumn{2}{c}{-} & \multicolumn{1}{c}{-} & \multicolumn{1}{c}{-} & \multicolumn{2}{| c}{-} & \multicolumn{1}{c}{-} & \multicolumn{1}{c}{-} \\
\hline
10 & -142&.15 & (25,1) & -142&.18 & 0.30 & (32,19,1) & \multicolumn{2}{| c}{-} & \multicolumn{1}{c}{-} & \multicolumn{1}{c}{-} \\
\hline
100 & -354&.0 & (26,0) & -354&.9 & 0.107 & (39,13) & -355&.2 & 0.107 & (47,21,10) \\
 & && & -354&.9 & 0.103 & (40,12) & -355&.1 & 0.108 & (46,22,10) \\
\hline
500 & -637&.8 & (26,0) & -645&.7 & 0.048 & (45,7) & -648&.1 & 0.048 & (58,16,4) \\
 & && & -645&.4 & 0.050 & (44,8) & -648&.0 & 0.050 & (57,16,5) \\
\hline
1000 & -810&.6 & (26,0) & -828&.8 & 0.035 & (47,5) & -834&.1 & 0.035 & (62,13,3) \\
 & && & -828&.4 & 0.034 & (48,4) & -834&.0 & 0.036 & (61,14,3) \\
\hline
2000 & -1021&.5 & (26,0) & -1061&.0 & 0.025 & (49,3) & -1073&.0 & 0.025 & (67,10,1) \\
 & && & -1056&.0 & 0.023 & (50,2) & -1072&.5 & 0.025 & (66,11,1) \\
\hline\hline
\end{tabular}
\end{table}

\begin{table}
\caption{Ground-state energies of ionized iron atoms over a range of
magnetic field strengths. Energies are given in units of keV\@. For
$B_{12} \ge 100$, the electron configuration of Fe atoms is such that
all of their electrons lie in the $\nu=0$ orbitals; therefore for
these field strengths the ionized atoms have all electrons in the
$\nu=0$ orbitals as well. The ionization state is designated by the
notation, ``Fe$^{n+}$,'' where $n$ is the number of electrons that
have been removed from the atom. The entry ``Fe$^{25+}$,'' for
example, is an iron nucleus plus one electron.}
\label{Fetable2}
\centering
\begin{tabular}{c | r@{}l | r@{}l | r@{}l | r@{}l | r@{}l | r@{}l | r@{}l | r@{}l | r@{}l | r@{}l}
\hline\hline
$B_{12}$ & \multicolumn{2}{| c}{Fe} & \multicolumn{2}{| c}{Fe$^{+}$} & \multicolumn{2}{| c}{Fe$^{2+}$} & \multicolumn{2}{| c}{Fe$^{3+}$} & \multicolumn{2}{| c}{Fe$^{4+}$} & \multicolumn{2}{| c}{Fe$^{5+}$} & \multicolumn{2}{| c}{Fe$^{10+}$} & \multicolumn{2}{| c}{Fe$^{15+}$} & \multicolumn{2}{| c}{Fe$^{20+}$} & \multicolumn{2}{| c}{Fe$^{25+}$} \\
\hline
100 & -354&.0 & -352&.8 & -351&.2 & -349&.0 & -346&.4 & -343&.2 & -318&.3 & -273&.8 & -199&.65 & -59&.01 \\
\hline
500 & -637&.8 & -635&.3 & -632&.0 & -627&.8 & -622&.7 & -616&.6 & -569&.4 & -486&.5 & -350&.2 & -99&.48 \\
\hline
1000 & -810&.6 & -807&.2 & -802&.8 & -797&.2 & -790&.7 & -782&.5 & -715&.8 & -602&.0 & -439&.6 & -122&.70 \\
\hline
2000 & -1021&.5 & -1016&.0 & -1008&.5 & -999&.8 & -989&.1 & -976&.7 & -905&.4 & -768&.6 & -546&.8 & -150&.10 \\
\hline\hline
\end{tabular}
\end{table}

\begin{table}
\caption{Fit of the ground-state energies of neutral and ionized iron
atoms and iron molecules to the scaling relation $E \propto
B_{12}^\beta$. The scaling exponent $\beta$ is fit over three magnetic
field ranges: $B_{12}=100-500$, $500-1000$, and $1000-2000$.}
\label{Fetable3}
\centering
\begin{tabular}{c | c c c c | c | c c}
\hline\hline
 & & & \multicolumn{5}{c}{$\beta$} \\
$B_{12}$ & Fe$^{25+}$ & Fe$^{20+}$ & Fe$^{10+}$ & Fe$^{+}$ & Fe & Fe$_2$ & Fe$_3$ \\
\hline
100-500 & 0.324 & 0.349 & 0.361 & 0.365 & 0.366 & 0.372 & 0.374 \\
\hline
500-1000 & 0.303 & 0.328 & 0.330 & 0.345 & 0.346 & 0.359 & 0.364 \\
\hline
1000-2000 & 0.291 & 0.315 & 0.339 & 0.332 & 0.334 & 0.358 & 0.363 \\
\hline\hline
\end{tabular}
\end{table}

\section{Conclusions}

We have presented density-functional-theory calculations of the
ground-state energies of various atoms and molecular chains (H$_N$ up
to $N=10$, He$_N$ up to $N=8$, C$_N$ up to $N=5$, and Fe$_N$ up to
$N=3$) in strong magnetic fields ranging from $B=10^{12}$~G to
$2\times10^{15}$~G\@.  These atoms and molecules may be present in the
surface layers of magnetized neutron stars, such as radio pulsars and
magnetars.  While previous results (based on Hartree-Fock or
density-functional-theory calculations) are available for some small
molecules at selected field strengths (e.g.,
Refs.~\citep{lai92,lai01,demeur94}) many other systems (e.g., larger C
molecules and Fe molecules) are also computed in this paper. We have
made an effort to present our numerical results systematically,
including fitting formulae for the $B$-dependence of the
energies. Comparison with previous results (when available) show that
our density-functional calculations tend to overestimate the binding
energy $|E_N|$ by about $10\%$. Since it is advantageous to use the
density functional theory to study systems containing large number of
electrons (e.g., condensed matter; see Ref.~\citep{medin06a}), it
would be useful to find ways to improve upon this accuracy.

At $B_{12} \ge 1$, hydrogen, helium, and carbon molecules are all more
energetically favorable than their atomic counterparts (although
for carbon, the relative binding between the atom and molecule is rather
small), but iron is not. Iron molecules start to become bound
at $B_{12}\agt 10$, and are not decidedly more favorable than isolated
atoms until about $B_{12}=100$.

For the bound molecules considered here, the ground-state energy per
atom approaches an asymptotic value as $N$ gets large. The molecule
then essentially becomes a one-dimensional infinite chain.
We will study such condensed matter in our companion paper \citep{medin06a}.

\begin{acknowledgments}
We thank Neil Ashcroft for useful discussion. This work has been
supported in part by NSF Grant No. AST 0307252, NASA Grant No. NAG 5-12034 and
{\it Chandra} Grant No. TM6-7004X (Smithsonian Astrophysical Observatory).
\end{acknowledgments}

\appendix

\section{Numerical method}

\subsection{Evaluating the integrals in the Kohn-Sham equations}

The two most computation-intensive terms in the Kohn-Sham equations
[Eq.~(\ref{kohneq})] are the ion-electron interaction term and the
direct electron-electron interaction term:
\be
V_{Ze,m}(z) =
\int d\mathbf{r}_\perp \, \frac{|W_m|^2(\rho)}{|\mathbf{r}-\mathbf{z}_j|}
\label{Vzemeq}
\ee
and
\be
V_{ee,m}(z) = \int \!\! \int d\mathbf{r}_\perp\,d\mathbf{r}' \,
\frac{|W_m|^2(\rho)\, n(\mathbf{r}')}{|\mathbf{r} - \mathbf{r}'|} \,.
\label{Veemeq}
\ee
Equation~(\ref{Vzemeq}), together with the exchange-correlation term,
$\int d\mathbf{r}_\perp \, |W_m|^2(\rho)\, \mu_{\rm exc}(n)$, can be
integrated by a standard quadrature algorithm, such as Romberg
integration \citep{press92}. Equation~(\ref{Veemeq}), however, is more
complicated and its evaluation is the rate-limiting step in the entire
energy calculation. The integral is over four variables ($\rho$,
$\rho'$, $z'$, and $\phi$ or $\phi-\phi'$), so it requires some
simplification to become tractable. To simplify the integral we use
the identity (see, e.g., Ref.~\citep{jackson98})
\be
\frac{1}{|\mathbf{r} - \mathbf{r}'|} =
\sum_{n=-\infty}^\infty \int_0^\infty dq \,
e^{in(\phi-\phi')} J_n(q\rho) J_n(q\rho') e^{-q|z-z'|} \,,
\ee
where $J_n(z)$ is the $n$th order Bessel function of the first kind. Then
\ba
V_{ee}(\mathbf{r}) & = & \int d\mathbf{r}' \,
\frac{n(\mathbf{r}')}{|\mathbf{r} - \mathbf{r}'|} \nonumber\\
 & = & 2\pi \int_{-\infty}^\infty dz' \int_0^\infty dq \,
J_0(q\rho) \left[\int_0^\infty \rho'\,d\rho' \, n(\rho',z') J_0(q\rho')\right]
\exp(-q|z-z'|) \,,
\ea
and
\ba
V_{ee,m}(z) & = & \int d\mathbf{r}_\perp \,
|W_m|^2(\rho)\, V_{ee}(\mathbf{r}) \nonumber\\
 & = & 4\pi^2 \int_{-\infty}^\infty dz' \int_0^\infty dq \,
\left[\int_0^\infty \rho\,d\rho \, |W_m|^2(\rho)\, J_0(q\rho)\right]
\left[\int_0^\infty \rho'\,d\rho' \, n(\rho',z') J_0(q\rho')\right]
\exp(-q|z-z'|) \,. \nonumber\\
\label{Veereq}
\ea
Using Eq.~(\ref{densityeq}) for the electron density distribution,
Eq.~(\ref{Veereq}) becomes
\be
V_{ee,m}(z) = \sum_{m'\nu'} \int_{-\infty}^\infty dz' \,
|f_{m'\nu'}(z')|^2 \int_0^\infty dq \, G_m(q) G_{m'}(q) \exp(-q|z-z'|) \,,
\label{Veezeq}
\ee
where
\ba
G_m(q) & = & 2\pi \int_0^\infty \rho\,d\rho \, |W_m|^2(\rho)\, J_0(q\rho) \nonumber\\
 & = & \exp(-q^2/2) L_m(q^2/2) \,,
\ea
and
\be
L_m(x) = \frac{e^x}{m!}\frac{d^m}{dx^m}(x^m e^{-x})
\ee
is the Laguerre polynomial of order $m$. These polynomials can be
calculated using the recurrence relation
\be
m L_m(x) = (2m-1-x) L_{m-1}(x) - (m-1) L_{m-2}(x) \,,
\ee
with $L_0(x)=1$ and $L_1(x)=1-x$.

Using the method outlined above the original four-dimensional integral
in Eq.~(\ref{Veemeq}) reduces to a two-dimensional integral. Once a
value for $z$ is specified, the integral can be evaluated using a
quadrature algorithm (such as the Romberg integration method).

\subsection{Solving the differential equations and total energy}

The Kohn-Sham equations [Eq.~(\ref{kohneq})] are solved on a grid in
$z$. Because of symmetry we only need to consider $z\ge0$, with $z=0$
at the center of the molecule. The number and spacing of the $z$ grid
points determine how accurately the equations can be solved. In this
paper we have attempted to calculate ground-state energies to better
than $0.1\%$ numerical accuracy. This requires approximately
(depending on $Z$ and $B$) 133 grid points for a single atom
calculation, plus 66 more for each additional atom in the molecule, or
a total of $\approx 66*(N+1)$ points for an $N$-atom molecule. The
grid spacing is chosen to be constant from the center out to the
outermost ion, then exponentially increasing as the potential decays
to zero. The maximum $z$ value for the grid is chosen such that the
amplitude of all of the electron wave functions $f_{m\nu}$ at that
point is less than $5\times10^{-3}$.

For integration with respect to $\rho$, $\rho'$, or $q$ (e.g., when
calculating the direct electron-electron interaction term), our
$0.1\%$-accuracy goal for the energy values requires an accuracy of
approximately $10^{-5}$ in the integral. A variable-step-size
integration routine is used for each such integral, where the number
of points in the integration grid is increased until the error in the
integration is within the desired accuracy.

Solving the Kohn-Sham equations requires two boundary conditions for
each $(m\nu)$ orbital. The first is that $f_{m\nu}(z)$ vanishes
exponentially for large $z$. Because the $f_{m\nu}(z)$ wave functions
must be symmetric or anti-symmetric about the center of the molecule,
there is a second boundary condition: wave functions with an even
number of nodes have an extremum at the center and wave functions with
an odd number of nodes have a node at the center; i.e.,
$f'_{m\nu}(0)=0$ for even $\nu$ and $f_{m\nu}(0)=0$ for odd $\nu$. In
practice, we integrate Eq.~(\ref{kohneq}) from the large-$z$ edge of
the $z$ grid and ``shoot'' toward $z=0$, adjusting $\varepsilon_{m\nu}$
until the boundary condition at the center is satisfied. One final
step must be taken to ensure that we have obtained the desired energy
and wave function shape, which is to count the number of nodes in the
wave function. For each $(m\nu)$ orbital there is only one wave function
shape that satisfies the required boundary conditions and has the
correct number of nodes $\nu$ (e.g., the shape of each orbital in
Fig.~\ref{wfCfig}, however complicated-looking, is uniquely
determined).

To determine the electronic structure of an atom or molecule
self-consistently, a trial set of wave functions is first used to
calculate the potential as a function of $z$, and that potential is
used to calculate a new set of wave functions. These new wave
functions are then used to find a new potential, and the process is
repeated until consistency is reached. In practice, we find that
$f_{m\nu}(z)=0$ works well as the trial wave function, and rapid
convergence can be achieved: four or five iterations for atoms and no
more than 20 iterations for the largest and most complex molecules. To
prevent overcorrection from one iteration to the next, the actual
potential used for each iteration is a combination of the
newly-generated potential and the old potential from the previous
iteration (the weighting used is roughly $30\%$ old, $70\%$ new).

\section{Correlation energy}

As was mentioned in Sec.~III, the form of the correlation energy has
very little effect on the relative energy between atom and molecule
(or between different states of the same molecule). This holds true
even if the calculations are done in the extreme case where the
correlation energy term is set to zero. As an example, consider the
energy of the C$_2$ molecule at $B = 10^{15}$~G\@. Using the
correlation energy of Skudlarski and Vignale [Eq.~(\ref{correq})], we
find the C atom has energy $E_a=-41\,330$~eV and the C$_2$ molecule has
energy per atom $E_m=-50\,760$~eV, so that the relative energy is
$\Delta E=9430$~eV\@. Using the correlation energy of Jones
[Eq.~(\ref{Joneseq})], we find $E_a=-44\,420$~eV and $E_m=-53\,840$~eV, so
that $\Delta E=9420$~eV\@. Without any correlation term at all,
$E_a=-38\,600$~eV and $E_m=-47\,960$~eV, so that $\Delta E=9360$~eV\@. As
another example of the relative unimportance of the correlation term,
two other works using density-functional calculations,
\citet{jones85,relovsky96}, find very similar cohesive energy (i.e.,
infinite chain) results even though they use two very different
correlation energy terms. For example, at $B = 5\times10^{12}$~G they
both find a cohesive energy of 220~eV for He$_\infty$.

We make one final comment about the accuracy of our chosen correlation
energy term, the Skudlarski-Vignale expression
Eq.~(\ref{correq}). \citet{jones85} found an empirical expression for
the correlation energy at high $B$ [Eq.~(\ref{Joneseq})], and then
checked its accuracy using the fact that the self-interaction of an
occupied, self-consistent orbital should be zero, i.e.,
\be
E_{\rm dir}[n_{m\nu}]+E_{\rm exc}[n_{m\nu}]=0 \,,
\label{selfeq}
\ee
where $n_{m\nu} = |\Psi_{m\nu}(\mathbf{r})|^2$ is the number density
of electrons in the $(m\nu)$ orbital. Performing such a test on the
Skudlarski-Vignale expression, we find that the error in
Eq.~(\ref{selfeq}) is of order $5$--$20\%$ for $B_{12}=1$ and up to
$20$--$30\%$ for $B_{12}=1000$ for the elements and molecules
considered here. Testing Jones's expression, we find it does as well
and sometimes better at $B_{12}=1$, but at large fields it does
considerably worse, up to $60$--$100\%$ error for $B_{12}=1000$. For
example, for He$_2$ at $B_{12}=1000$ the Skudlarski-Vignale
correlation function satisfies Eq.~(\ref{selfeq}) to within $23\%$ but
Jones's expression satisfies Eq.~(\ref{selfeq}) only to within
$63\%$. Thus, the Skudlarski-Vignale correlation function adopted in
this paper is much more accurate than Jones's expression for a wide
range of field strengths.

\end{document}